\documentclass[a4paper]{llncs}

\usepackage{amsmath,amsthm,amssymb}
\usepackage{mathtext}
\usepackage[utf8]{inputenc}
\usepackage[english]{babel}

\usepackage{natbib}
\usepackage{xcolor}

\usepackage{listings}
\usepackage{color}

\definecolor{mygreen}{rgb}{0,0.6,0}
\definecolor{mygray}{rgb}{0.5,0.5,0.5}
\definecolor{mymauve}{rgb}{0.58,0,0.82}

\usepackage{geometry}
\usepackage{graphicx}
\usepackage{subcaption}
\usepackage{mathtools,hyperref}
\usepackage{footnote}
\usepackage{enumitem}
\usepackage[boxed, algosection, linesnumbered, titlenumbered]{algorithm2e}

\newtheorem{Stat}{Statement}
\usepackage{array} 
\renewcommand{\eqref}[1]{(\ref{#1})}                
\newcommand{\figref}[1]{Figure \ref{#1}}            
\newcommand{\tblref}[1]{Table \ref{#1}}             
\newcommand{\algref}[1]{Algorithm \ref{#1}}         

\DeclareUnicodeCharacter{03C0}{$\pi$}

\begin{document}

\title{Reliability of Checking an Answer Given
by a Mathematical Expression in Interactive Learning Systems}

\author{    Vladimir G. Danilov         \and
            Ilya S. Turuntaev
}

\institute{
    National Research University Higher School of Economics,\\ 
    vgdanilov@mail.ru \\
    isturunt@gmail.com
}

\maketitle

\begin{abstract}

In this article we address the problem of automatic answer checking
in interactive learning systems that support mathematical notation.
This problem consists of the problem of establishing identities in 
formal mathematical systems and hence is formally unsolvable.
However, there is a way to cope with the issue.
We suggest to reinforce the standard algorithm for function comparison
with an additional pointwise checking procedure.
An error might appear in this case.
The article provides a detailed analysis of the probability of this error.
It appears that the error probability is extremely low in most common cases.
Generally speaking, this means that such an additional checking procedure
can be quite successfully used in order to support standard algorithms for
functions comparison.
The results, obtained in this article, help avoiding some sudden effects of
the identity problem, and provide a way to estimate the reliability of answer
checking procedure in interactive learning systems.

\end{abstract}

\section*{Introduction}

In this article, we propose a solution for the problem of establishing identities in formal
mathematical systems. This problem is known to be formally unsolvable (in fact, this is a corollary
of the G\"odel's theorem). And the problem may appear even in computer systems
supporting analytic calculations (see an example below). We consider this problem in connection with the development
of an interactive educational system where a computer algebra system is used
as a computational module and the nonsolvability of the identity problem may result
in an inappropriate behavior.
A more detailed overview of our interactive educational system
is given further in the article. Here we will just point out
that the system generates training tasks and performs answer checking,
where the human-obtained answer in the form of a mathematical expression
is compared with the one obtained by the system.
This is where the problem of identity checking appears.
In this article, we suggest to reinforce (if necessary) the native algorithm
for comparing mathematical expressions by the computer algebra system
by an additional pointwise checking procedure.
The article provides a detailed investigation of the probability of error
which appear in this case.

Roughly speaking, it appears that this probability is much less than 
any reasonable value.
For instance, this value could be of the order of the probability
of obtaining a non-unique result when shuffling a deck of $ 36 $ cards (which happens to be
about $ 0.27 \cdot 10^{-41} $).
More precisely, this is true if we exclude some exotic cases like
frequently oscillating functions.

We end up with the following scheme for checking the answer:
\begin{itemize}
    \item compare two answers using computer algebra system's standard tools;
    \item if the result is indefinite, run the pointwise checking procedure
    in order to improve the result.
\end{itemize}
The article is divided into two main sections.
The first section provides a brief overview of the interactive educational system
and the problem of automatic answer checking.
In Section 2, we present an algorithm for additional answer checking and
provide a detailed analysis of the algorithm error probability.

\section*{Interactive Educational System Overview}

There are plenty of approaches to the e-Learning systems development:
from simple databases of classified theoretical material
to complex interactive intelligent systems. Lots of achievements had been
reached in the delivery of educational courses;
one can say that the key idea of democratic education is now being embodied.
In this article, we will consider the problem of training tasks organization.
Consider classes on, say, Mathematical Analysis at some first grades in higher school.
Students face a number of training tasks to work the appropriate theoretical material out.
The main sources for such tasks are schoolbooks.
Let us call it the traditional approach to the organization of practical classes.
This approach has some major drawbacks:

\begin{itemize}
    \item
        Following this approach, one cannot take into account the individual characteristics
        of each student in a group. This means that the professor provides tasks
        according to some concept of an average student. But in fact each student needs
        his own individual amount of tasks to succeed in learning something. While one has
        to work out say 5 tasks to ensure the complete understanding of the material,
        another student needs 15 tasks to reach the same result. And this is not up to
        the professor to develop some individual working plans for each student in a group.
        This requires too much time and will heavily increase the demands.
    \item
        When talking about some kind of home tasks, professors may spend too much
        time just to check all tasks up. This is a mechanical job but it needs to be done.
    \item
        It is nearly impossible to create a unique amount of tasks for each student.
        No book is large enough to provide that amount of different but at the same time
        equivalent tasks.
    \item
        Student may suffer a lack of information if he encounters an error.
        The thing is that if you fail you will not get any additional information
        about that particular problem from the schoolbook.
\end{itemize}

We claim that the situation can be improved.
The main idea here is that the practical part of educational process, and especially
training tasks generation, delivery and processing, can be highly automatized.
We develop this approach in our Interactive Educational System --- a project
for educational process support where we aim to improve educational process
for exact sciences to make it more effective and less painful.

Our Interactive Educational System is a web application, where we concentrate
on solving the problems of task creation and delivery and answer checking providing
a high level of automatization.
The approach of training task automatization is based
on the following three general concepts:

\begin{itemize}
    \item task templates
    \item random generation
    \item automatic answer checking
\end{itemize}

The concept of task templates prescribes to create some general definitions
for some task classes instead of defining each task separately.
Such task templates include a number of indefinite parameters,
where each parameter takes its values from its own set.
These parameters, being defined appropriately, define a concrete task from
the task class defined by the template.
An important thing to point out is that all concrete tasks that can be retrieved
from some task template are equivalent in a sense meant by the author.
In other words, it is up to the task author to decide
what equivalent tasks should look like.
In fact, he makes this decision naturally, behind the scenes, each time he creates
a new task template.

To define a task, one needs to specify its problem (the problem formulation), solution
(a sequence of steps with some explanations that describes how the task is solved),
and answer (which in our case is a mathematical expression).
Similarly, to define a task template,
one needs to specify its problem, solution, and answer templates.
Let us take a look at the problem formulation.
It usually includes some few sentences that describe
what needs to be done and some special objects.
In the case of exact sciences,
these objects are usually some mathematical expressions ---
here and further we consider precisely this case.
So problem formulation consists of some natural-language sentences
and some mathematical expressions.
The situation is similar to the solution,
while an answer must be a strict object, i.e., a mathematical expression.
In the Interactive Educational System, we use Markdown\footnote{
    Markdown is a markup language by John Gruber with plain-text formatting syntax which is
    easy to read and manage. Official website: \url{http://daringfireball.net/projects/markdown/}
} language for the text markup and our special language
for defining the task dynamic attributes --- SmallTask.
The SmallTask language was designed specifically to serve the needs of task template definition
and provides tools for mathematical expressions declaration, evaluation, and representation.

Given these tools and a special editor, a user can define task templates
which the system can further concretize
providing additional calculations automatically.
Let us take a trivial problem from Mathematical Analysis class.
Consider a task on the chain rule.
If one wants to teach somebody this rule, he will probably explain it to the student
and provide him with some special tasks on the subject.
The problem formulation for such task may look as follows:

\begin{minipage}[c]{0.5\textwidth}
    \textit{Calculate the derivative of} \( \sin(\cos(x)) \).
\end{minipage}

The solution explanation for this problem may look as follows:

\begin{minipage}[c]{0.5\textwidth}
    \begin{itemize}
        \item According to the chain rule
        \(\frac{d }{dx} \sin(\cos(x)) = \frac{d\sin(y)}{dy}|_{y=\cos(x)} \cdot \frac{d\cos(x)}{dx}\)
        \item \(\frac{d \sin(y)}{d y} = \cos(y)\)
        \item \( \frac{ d \cos(x) }{d x} = - \sin(x) \)
        \item Combining the above results together we get:
            \[
                \frac{d }{dx} \sin(\cos(x)) = \cos(y)|_{y=\cos(x)} \cdot (- \sin(x)) = - \cos(\cos(x)) \cdot \sin(x)
            \]
    \end{itemize}
\end{minipage}

The answer to the problem is \( - \sin(x) \cdot \cos(\cos(x))\).

These three declarations (problem, solution, and answer) together
define one training task for the chain rule.
In fact, to create a new task, we can just replace sine and cosine with some other functions
in the problem formulation and perform appropriate changes in solution
and answer.
One may define the general problem formulation for all these tasks:

\begin{minipage}[|c|]{0.5\textwidth}
    \textit{Calculate the derivative of} \( f(g(x)) \).
\end{minipage}

Here $f$ and $g$ belong to some function class (i.e., of base elementary functions).
Performing appropriate changes in the solution and answer declarations,
we end up with a generalization of training tasks on the chain rule
which, in fact, is the task template introduced earlier.
In the Interactive Educational System, we allow the users to define such task templates
and also to specify the way the undefined attributes (such as $f$ and $g$ in the example above)
should be defined. For instance, $f$ and $g$ in the example above may be some random
base elementary functions (such as $sin$, $cos$, $exp$, $log$, etc.).
Given such declaration, the system will automatically
replace all occurrences of such attributes
by appropriate values retrieved automatically
according to their randomization declaration.
That is what the random generation concept is about.
This is done each time someone calls a task: the system creates
the concrete task problem, solution, and answer according to the template
and then the user gets a concrete problem formulation
while the appropriate solution and answer are
stored on the server for further needs.
Given the problem formulation, a student tries to solve the problem and
after that he is to put his answer into a special field on the page
and to confirm it.
Then this answer is sent to the server where it is checked against
the answer calculated by the system in order to decide whether a student
succeeded or not. The appropriate correct solution is sent back to the student
in the latter case.

The most challenging thing here is the automatic answer checking.
At the same time, this is the key idea of the project.
Together with the templating system and randomization,
it forms a mechanism which simplifies the process
of training tasks creation, delivery, and checking,
which is a heavy and essential part of practical classes.

\section*{Identity Problem and the Probability of Pointwise Checking Procedure Error}

The problem of automatic answer checking is rigidly coupled with the problem of determining
whether two expressions are semantically equal. We call two mathematical expressions
\textit{semantically equal} if they both represent the same mathematical object.
The concept of semantic equality is opposite of the concept of \textit{syntactic equality},
which means that two expressions have exactly the same representation.
For example, $ 2^{x} $ and $ e^{x \log 2} $ are semantically equal,
since they both represent the same object, but they are syntactically different
by means of the given representation in mathematical notation.
It is obvious that the problem of testing the semantic equality of two expressions
is equivalent to the problem of checking whether a particular expression
equals zero (since $ f_1 \equiv f_2 $ if and only if $ g = f_1 - f_2 \equiv 0 $).
The latter is known as the \textit{Constant Problem} \footnote{
    It is also referred to as the \textit{Identity Problem}
}.
It was shown (see \cite{richardson1969some}, also \cite{laczkovich2003removal})
that this problem is undecidable for a class of functions that contains
$ log2 $, $ \pi $, $ e^x $, $ \sin(x) $ and $ |x| $.
This result is known as Richardson's theorem.

The undecidability of the Constant Problem raises a huge problem
in the learning system development.
Automatic answer checking is one of the general features of the system
but there is still no way to guarantee the correctness of the answer comparison
using computer algebra systems' native algorithms.
Although the problem may seem quite seldom or even exotic,
it needs to be examined, since its side effects can appear unexpectedly.
In our investigation, we have discovered that one of the tested
computer algebra systems (CAS) was unable
to determine whether $ 2^x $ equals $ e^{x \cdot \log(2)} $ or not.
That exact problem was fixed in further versions of that CAS,
but the general problem still remains open.
In order to cope with the issue, we have to impose some limitations
on the answers in the system
and to provide an algorithm for additional answer comparison,
which should increase the probability of correct answer checking.

For most of CAS, the undecidability of the constant problem
results in the situation where, in some cases,
a system is unable to make a decision
whether the two given expressions are equal or not, thus
returning nothing instead of \textit{True} or \textit{False}.
Such result must be treated as the ``I don't know''-answer,
which means that we are unable to say for sure whether the user's answer is correct or not.
But still we need to make a decision.
There are two trivial ways to do this:
always consider such an ambiguous answer as correct or, on the contrary,
always consider such an answer as incorrect.
The second alternative may lead to a critical misunderstanding
and it also violates the student's rights in a way, namely,
one can say the student is unfairly punished.
It may force bad ratings among the students, and this is definitely unacceptable.
From this point of view,
the first alternative seems a bit less harmful, namely,
no one is directly hurt because of the answer checking error ---
most of the time it is not detected at all.
However, this is definitely not the best way to solve the problem.
A better way is to provide an algorithm that performs additional tests on answers
if the CAS is unable to give an acceptable result.

We suggest the following algorithm for additional answer comparison,
which runs if it is impossible to determine the answer correctness safely,
i.e., the CAS answer is uncertain.
Consider task $ T $ with an answer that is a mathematical function of a single variable.
Let $ f_{real}(x) $ be the real answer to $ T $ problem obtained by the system,
and let $ f_{user}(x) $ be a student-calculated answer.
The algorithm accepts $ f(x) = f_{real}(x) - f_{user}(x) $
and tries to determine whether the equality $ f(x) \equiv 0 $ is correct.
Here is its pseudocode:

\begin{algorithm}[H]
\SetKw{Continue}{continue}
\KwData{
    $f$ --- the function to be tested, \\
    $m$ --- a natural number,\\
    $A, B$ --- real numbers ($ A < B $)}
\KwResult{$False$ or $True$}
rnd = Rnd(A, B)\;
used = set()\;
$f$ = limitExecTime($f$, $\Delta$)\;
\For{$ i \in \{0..m-1\}$}{
    d = rnd.next()\;
    \While{d $\in$ used}{
        d = rnd.next()\;
    }
    used.add(d)\;
        tmp = $ f $(d)\;
        \If{tmp is $ None $}{
            $ i-- $\;
            \Continue\;
        }
    \If{tmp != 0}{
        \Return{$False$}\;
    }
}
\Return{$True$}\;
\caption{Basic algorithm for additional answers comparison}\label{alg:add-compare-pseudo}
\end{algorithm}

The notion of the algorithm is quite simple:
it repeatedly picks up a random number $ d $ from $ \mathcal{U}(A, B) $
until either the maximum number of iterations $ m $ is reached
or $ f(d) \neq 0 $, i.e., a non-zero point is found\footnote{
    $ \mathcal{U}(A,B) $ is a uniform distribution on $ [A, B] $.
}.
The $ True $ value is returned in the first case
(which corresponds to the positive result of comparison,
i.e. $ f_{real} = f_{user} $).
In the second case, the algorithm returns $ False $ which means that
$ f = f_{real} - f_{user} \not\equiv  0$.
There is a special limitation on the selected points:
we require them to be different, since there is no need to test the function value twice.
For that purpose,
a set of used points (called $ used $ in the pseudocode above) is stored;
on each iteration, we demand that the new point does not belong to this set,
and then we populate the set with this new point.

Although the algorithm suggested above is rather simple,
it has several advantages.
The first thing to mention is that it is ``student-friendly'':
the only case the algorithm fails is when it returns $ True $,
while, in fact, the student's answer is incorrect and,
as this will be shown below, this situation is quite seldom.
Another thing to notice is that we limit execution time for $ f $.
The function value computation can be really slow in some definite cases,
while we do not really care about each and every concrete value ---
the only thing we need is a number of points to test the function value at.
So if the computations run out of some predefined time limit,
we just pick another point instead of that ``complicated'' point.

The result of \algref{alg:add-compare-pseudo} is not exact.
In fact, it can fail if the $ f_{real} $ and $ f_{user} $ functions
are too close to each other and the maximum number of points to check is too low.
In order to estimate its error probability, the following proof is suggested.

\subsection*{Analysis of Additional Answer Checking Algorithm}

Consider $ f = f_{real} - f_{user} $, which is a real-valued elementary function.
Let $ A, B \in \mathbb{R} $ be such that $ A < B $ and $ f $ is an analytic function
on $ [A, B] $. We use \algref{alg:add-compare-pseudo} to determine
whether $ f \equiv 0 $ or not.

\begin{Stat}
    If actually $ f \equiv 0 $, then \algref{alg:add-compare-pseudo} always
    returns the correct result.
\end{Stat}

Indeed, if $ f \equiv 0 $, then the statement $ f(d) = 0 $ is correct
for every $ d $ picked on each iteration of \algref{alg:add-compare-pseudo} main loop,
and thus the program successfully passes through all loop iterations and returns
$ True $ at the end.

Now let $ f \not\equiv 0 $, and let $ k \in  \mathbb{N} $ be such that\footnote{
    By $ |M| $ we denote the cardinality of the set $ M $.
}
$ \left|\{x \in [A, B] : \quad  f(x) = 0\}\right| < k$.
In other words, $ k $ gives an upper estimate for the number of elements
at the zero locus of $ f $.
And finally, let $ m $ be the maximum number of check points
(as in \algref{alg:add-compare-pseudo}).
There are the following two main possibilities:

\begin{enumerate}
    \item $ k < m$
        --- the number of points to be checked is greater than
        the maximum number of possible zeros of $ f $;
    \item $ k \geq m $
        --- it is possible that $ f $ has more zeros on $ [A, B] $ than the number of points
        checked in \algref{alg:add-compare-pseudo}.
\end{enumerate}

\begin{Stat}
    If $ k < m $, then \algref{alg:add-compare-pseudo} always returns the correct result.
\end{Stat}

\begin{proof}
    The only way for the algorithm to fail is occasionally to pick every $ d $ such that
    $ f(d) = 0 $. Let $ d_0, d_1, \dots, d_{m-1} $ be a random sample chosen by the
    algorithm. Because of the limitation discussed above, all $ d_i $ ($ i \in 0..m-1$)
    are different: $ d_0 \neq d_1 \neq \dots \neq d_{m-1} $.
    Consider the first $ k $ iterations of the \algref{alg:add-compare-pseudo} main loop.
    The worst is here the case when $ \forall i \in 0..k-1 $ $ f(d_i) = 0 $,
    in other words, the first $ k $ points were accidentally chosen from the zero locus of $ f $
    (otherwise, the algorithm would have returned $ False $ at some iteration).
    Now at the $ (k+1) $ iteration, the algorithm necessarily picks $ d_k $
    such that $ f(d_k) \neq 0 $.
    Indeed, if $ f(d_k) = 0 $, then either $ \exists j \in 0..k-1 $ such that $ d_k = d_j $,
    which is forbidden by our limitation, or $ f $ has more than $ k $ zeros, which
    contradicts the definition of $ k $.

    And so the algorithm will always return $ False $ (in $ (k+1) $ steps in the worst case).
\end{proof}

So there is only one possibility left for \algref{alg:add-compare-pseudo} to fail: $ k \geq m $.
In this case, it is possible that we get $ f(d_i) = 0 $
for each $ d_i $ ($ \forall i \in 0..m-1 $).
In order to estimate the algorithm correctness,
we analyze the probability of such a situation.

Let $ Z = \{x \in [A, B]:\,\, f(x) = 0\} $, and let $ k = |Z| \geq m $.
The probability of \algref{alg:add-compare-pseudo} failure
is exactly the probability of picking $ d $ from $ Z $ one by one $ m $ times in a row.
When $ [A,B] $ is a continuous closed line segment
and $ d $ has a continuous uniform distribution on $ [A,B] $,
then $ \mathbb{P}(d_i \in Z) = 0$, since $ Z $ is a discrete subset of the continuous set
$ [A, B] $.
But the situation here is a little bit different.
Talking about computer implementation of the suggested algorithm,
we must consider the $ [A, B] $ segment to be a discrete sequence of points
defined by the floating point numeral system we use.
From this point of view, we deal with a grid of floating point numbers
with spacing determined by an appropriate rounding procedure.
So let $ AB $ be a set of floating point numbers inside $ [A, B] $.
And let $ M = |AB|$.
Now the probability of \algref{alg:add-compare-pseudo} is determined
by the probability of picking $ m $ points one by one from a $ k $-subset
of a set of $ M $ elements.

The probability of choosing a point from a set $ AB $ of $ M $ elements
such that it belongs to its $ k $-subset $ Z $ is given
by \eqref{eq:choose-1-of-k-of-M}.

\begin{equation}
    \label{eq:choose-1-of-k-of-M}
    \mathbb{P}(d \in Z) = \frac{k}{M}
\end{equation}

When the first point has been picked and it appears to be from $ Z $,
we pick another one as prescribed by the algorithm.
Now in this situation (after we have already chosen the first point),
we choose the second one from a set of $ (M-1) $ elements,
and the ``zero-set'' now consists of $ (k-1) $ points.
So the probability of such an event is given
by the conditional probability \eqref{eq:choose-another-one}.

\begin{equation}
    \label{eq:choose-another-one}
    \mathbb{P}\left(d_1 \in Z | d_0 \in Z\right) = \frac{k-1}{M-1}
\end{equation}

The total probability of choosing two points in a row such that both of them belong to $ Z $
is given by \eqref{eq:choose-2-in-a-row}.

\begin{equation}
    \label{eq:choose-2-in-a-row}
    \mathbb{P}\left( d_1 \in Z, d_0 \in Z \right) =
        \mathbb{P} \left( d_0 \in Z \right) \cdot \mathbb{P} \left( d_1 \in Z | d_0 \in Z\right) =
        \frac{k \cdot (k-1)}{M \cdot (M-1)}
\end{equation}

If we continue the same reasoning, we will obtain formula \eqref{eq:choose-m-in-a-row}
for the total probability of choosing $ m $ different points in a row
from $ AB $ such that all of them belong to $ Z $.

\begin{equation}
    \label{eq:choose-m-in-a-row}
    p_{M,m,k} =
    \mathbb{P} \left(d_{m-1} \in Z, d_{m-1} \in Z, ..., d_0 \in Z \right) =
        \prod\limits_{i=0}^{m-1} \frac{k-i}{M-i} = \frac{k! (M-m)!}{M! (k-m)!}
\end{equation}

Formula \eqref{eq:choose-m-in-a-row} holds for $ m \leq k \leq M $.
In the case $ k > M $, we consider the failure probability equal to $1$.
So finally, the probability of \algref{alg:add-compare-pseudo} error
is given by the expression \eqref{eq:full-prob}.

\begin{equation}
    \label{eq:full-prob}
    P_{err} = \begin{cases}
        0,          & k < m, \\
        p_{M,m,k},  & m \leq k \leq M,\\
        1,          & k > M.
    \end{cases}
\end{equation}

Formula \eqref{eq:full-prob} depends on the following three parameters:
$ M $ (the number of points in the segment being examined),
$ m $ (the maximum number of points to be checked, $ m \leq M $),
and $ k $ (the upper bound to the number of zeros of $ f $).
Some estimates are required to retrieve the numeric value of the error probability.
Given some estimates for $ N $ and $ k $,
we can choose the $m$ values such that $ P_{err} $ is small in a certain sense.

\subsection*{Error Probability Analysis}

In machine arithmetics, we often deal with a grid of points which approximate real numbers
according to the rounding procedure of the appropriate floating point numerical system.
The spacing of such a grid changes at the perfect powers of the base of the system
and is non-decreasing with respect to the absolute value of a number.
One of the most general characteristics of a floating point numerical system
is its \textit{machine epsilon}, i.e.,
an upper bound on the relative error in the system.
This is often defined as the maximum $ \varepsilon $
such that $ 1 + \varepsilon $ is rounded to $ 1 $
(because the worst relative error appears when the rounding procedure
is applied to such numbers; more on the topic in \cite{goldberg1991every}).
So given a segment $ [A, B] $ and a floating point numerical system,
we can calculate all spacings inside $ [A, B] $
and thus retrieve the number of points (grid nodes) inside the segment.
But such calculations would be too complicated in our case,
so it is better to estimate that number somehow else.

Let $ \varepsilon_x > 0 $ be the maximum number such that $ |x| + \varepsilon_x $
is rounded to $ |x| $.
And let $ 0 < x_0 = A < x_1 < x_2 < \dots < x_{\tilde{M}-1} = B $
be the nodes of the floating point grid inside $ [A, B] $ ($ 0 < A < B$).
Then the following inequality \eqref{eq:A-B-spacings} is correct
since grid spacing is not decreasing.

\begin{equation}
    \label{eq:A-B-spacings}
    \varepsilon_A = \varepsilon_{x_0} \leq \varepsilon_{x_1} \leq ... \leq
        \varepsilon_{x_{\tilde{M}-1}} = \varepsilon_B
\end{equation}

Inequality \eqref{eq:A-B-spacings} implies inequality \eqref{eq:M-estimate}.
We take this $ M $ as an estimate for the quantity of floating point numbers inside $ [A, B] $,
and $ \tilde{M} $ is a real quantity.
This shows that $ [A,B] $ contains at least $ M $ points.

\begin{equation}
    \label{eq:M-estimate}
    M = \left[ \frac{B-A}{ \varepsilon_B} \right] \leq  \tilde{M}
\end{equation}

If $A < B < 0$, then a new interval $ [-B, -A] $ can be considered.
In the case $ A = 0 $ and $ B > 0 $, we may consider an estimate for $[\tilde{A}, B]$,
where $ \tilde{A} $ is such that $0 < \tilde{A} \leq x$
for any floating point number $x > 0$.

\begin{figure}[h]
    \centering
    \includegraphics[width=0.6\linewidth]{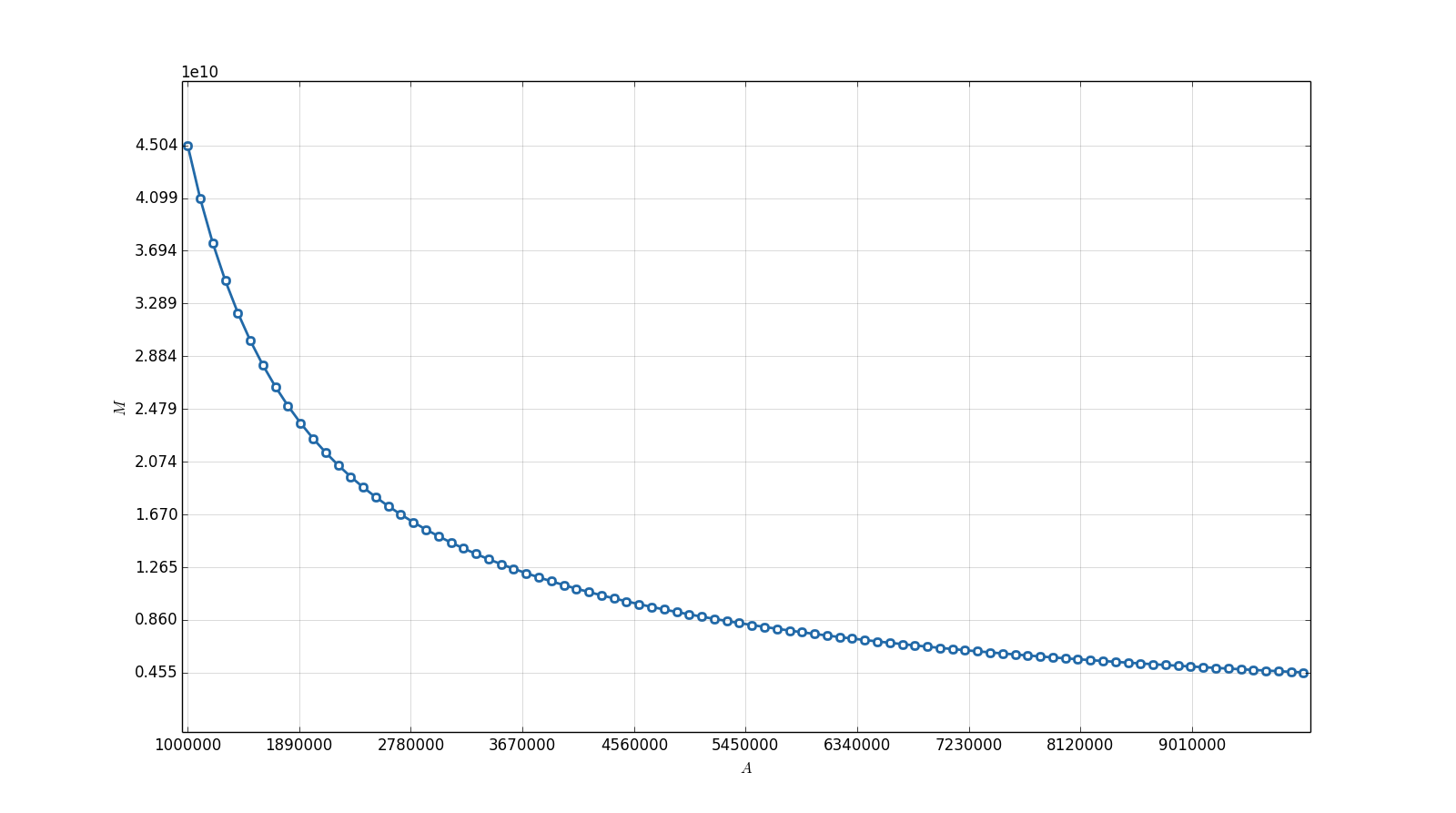}
    \caption{The estimate of the number of points in $ [A, A+5] $
    with respect to $ A $ (NumPy.float64) }
    \label{fig:M-wrt-A}
\end{figure}

In order to illustrate the values of $ M $,
an example for the \textbf{NumPy.float64} floating point data type\footnote{
    \textbf{NumPy} is a Python package for scientific computing.
    \textbf{NumPy.float64} is a double precision float data type provided by \textbf{NumPy}.
} is provided.
Some numeric values are provided in \tblref{tbl:M-wrt-A}.
The plot in \figref{fig:M-wrt-A} shows the value of $ M $ for segments
like $ [A, A+5] $ with respect to $ A $.

\begin{savenotes}
    \begin{table}
    \centering
        \begin{tabular}{| >{\centering\arraybackslash}m{2in} | >{\centering\arraybackslash}m{2in} |}
                        \hline
            \textbf{$ [A,A+5] $}        &       \textbf{$M$}        \\
            \hline\hline
                                    [10, 15]                        &   3002399751580330               \\
            \hline
                                    [100, 105]                      &   428914250225761                \\
            \hline
                                    [1000, 1005]                    &   44811936590751                 \\
            \hline
                                    [10000, 10005]                  &   4501348952894                  \\
            \hline
                                    [100000, 100005]                &   450337445864                   \\
            \hline
                                    [1000000, 1000005]              &   45035771094                    \\
            \hline
                                    [10000000, 10000005]            &   4503597375                     \\
            \hline
                                    [100000000, 100000005]          &   450359940                      \\
            \hline
                                    [1000000000, 1000000005]        &   45035996                       \\
            \hline
            
        \end{tabular}
    \caption{Values of $ M $ for segments like $ [A, A+5] $}
    \label{tbl:M-wrt-A}
    \end{table}
\end{savenotes}

\begin{figure}[h]
    \centering

    \begin{subfigure}[b]{0.4\textwidth}
        \includegraphics[width=\textwidth]{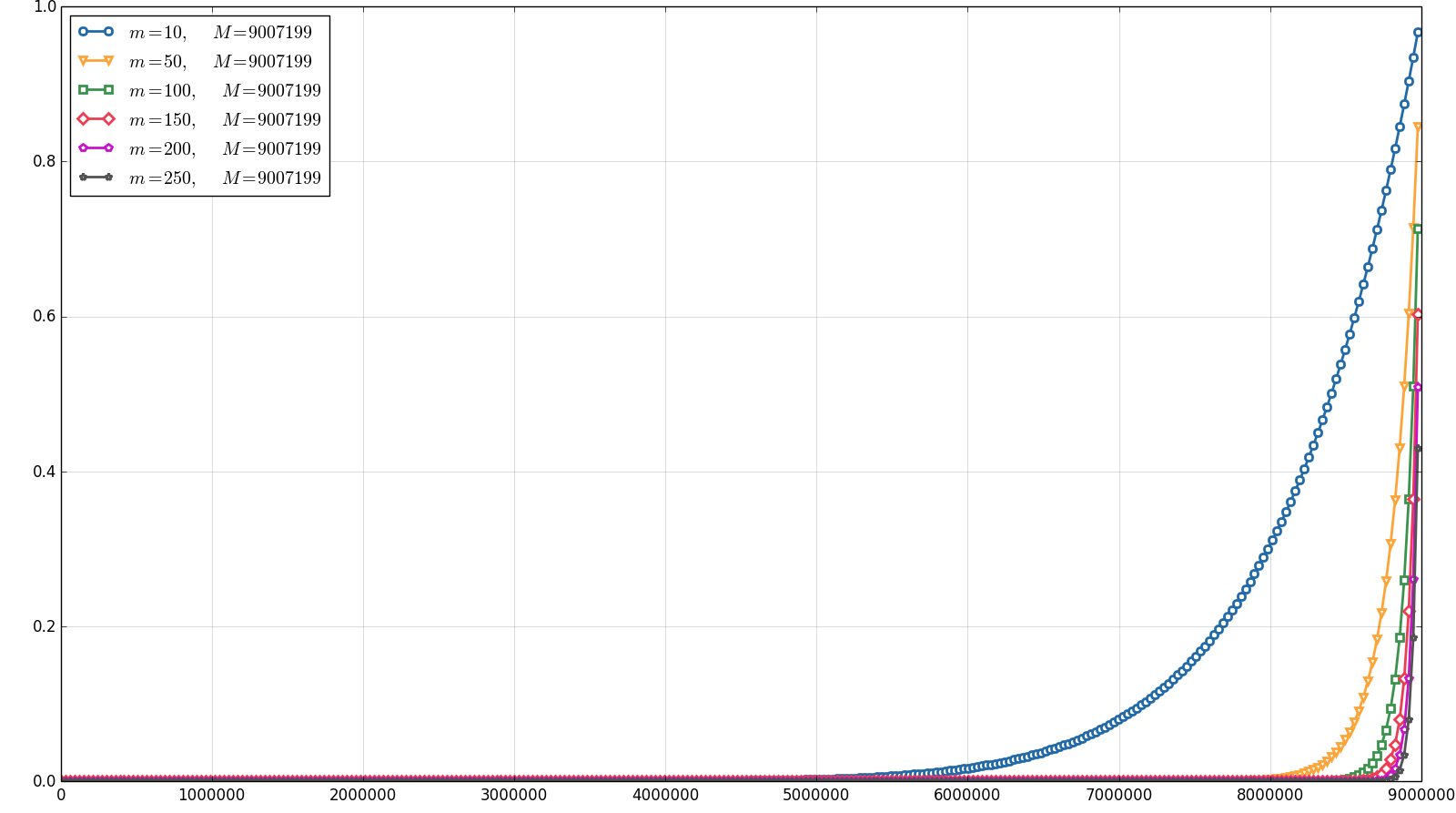}
        \caption{}
        \label{}
    \end{subfigure}
    ~
    \begin{subfigure}[b]{0.4\textwidth}
        \includegraphics[width=\textwidth]{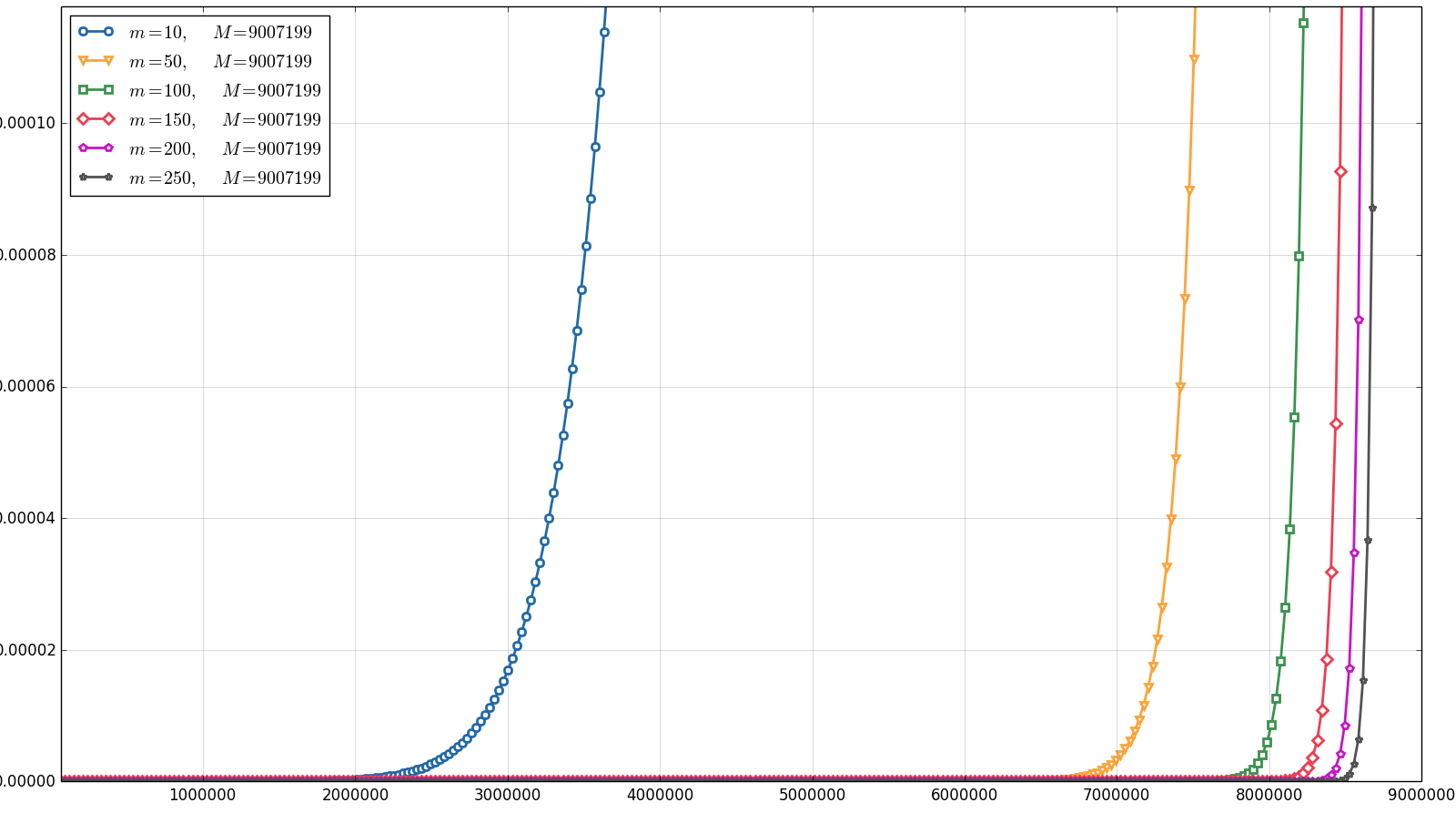}
        \caption{}
        \label{}
    \end{subfigure}

    \caption{$ p_{M,m,k} $ for $ [A, B] = [10^9, 10^9+1] $}
    \label{fig:pMmk-10e9}
\end{figure}

Looking at $ p_{M,m,k} $ given by \eqref{eq:choose-m-in-a-row},
one may notice that, given some fixed values of $ k $ and $ m $,
the value of $ p_{M,m,k} $ decreases with respect to increasing $ M $,
namely, the bigger the value of $ M $, the lower the probability.
And also the values of $ M $ are really big ---
so big that it is likely that, in most cases, either
the value of $ k $ is extremely lower or the examined function equals $0$.
However, we still need to introduce some estimates.
Some plots are provided in order to illustrate the error probability behavior
(\figref{fig:pMmk-10e9}).

\begin{figure}[h]
    \centering
    \begin{subfigure}[b]{0.4\textwidth}
        \includegraphics[width=\textwidth]{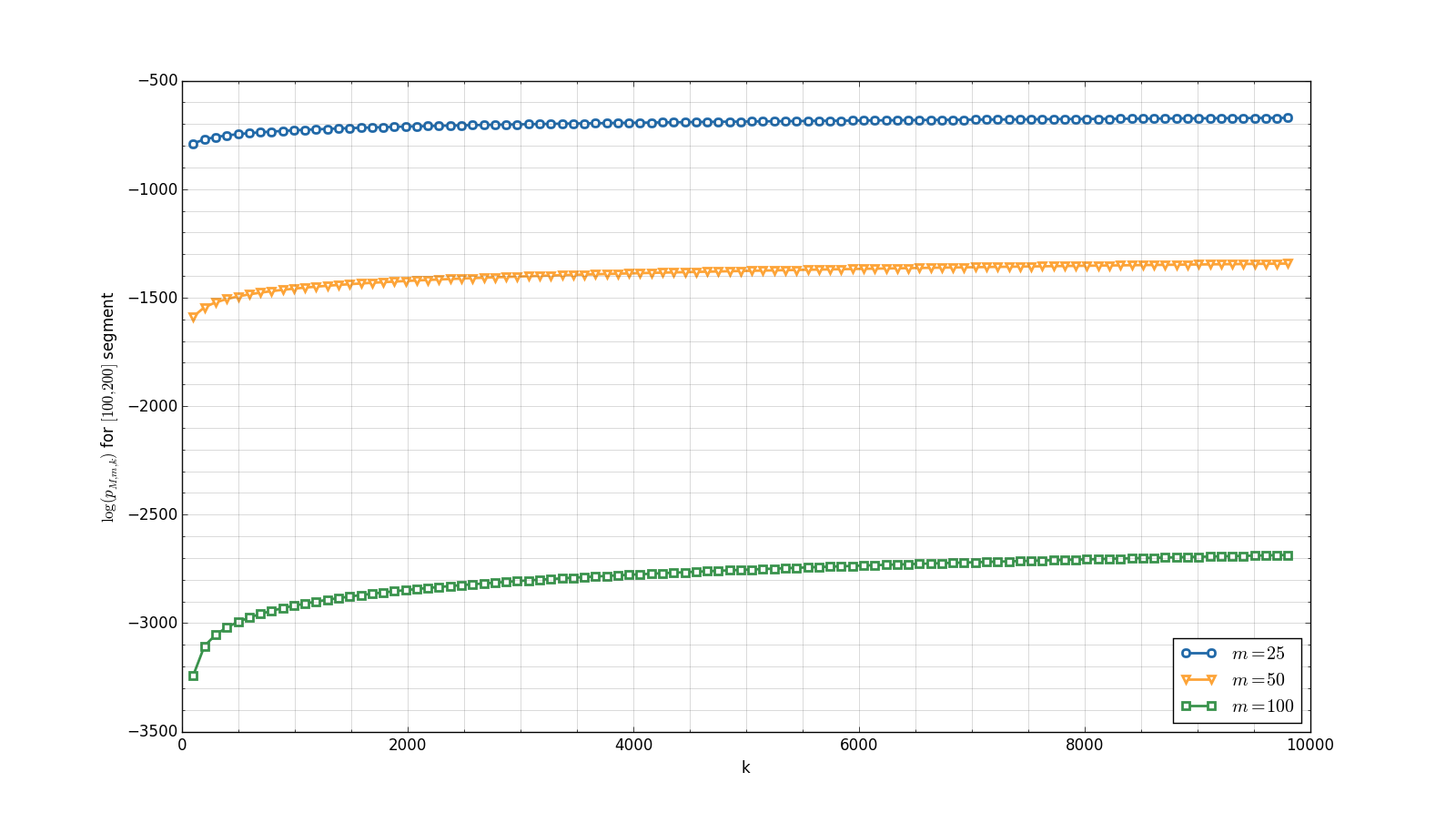}
        \caption{$\log(p_{M,m,k})(k)$ for $[100, 200]$ segment, $m \in \{25, 50, 100\}$}
        \label{}
    \end{subfigure}
\   ~
    \begin{subfigure}[b]{0.4\textwidth}
        \includegraphics[width=\textwidth]{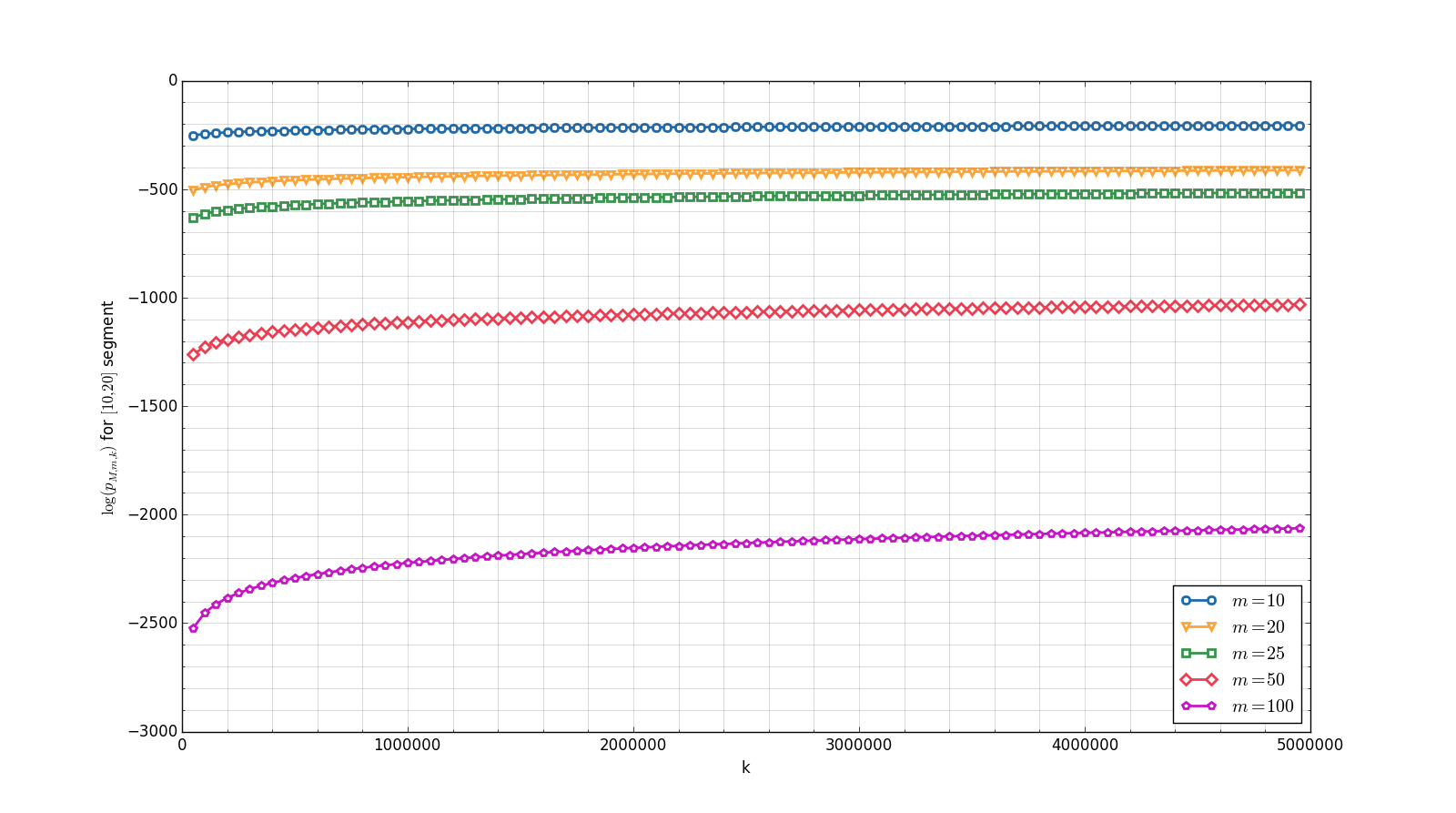}
        \caption{$\log(p_{M,m,k})(k)$ for $[10, 20]$ segment, $m \in \{10, 20, 25, 50, 100\}$}
        \label{fig:logp-10-20}
    \end{subfigure}
\\
    \begin{subfigure}[b]{0.4\textwidth}
        \includegraphics[width=\textwidth]{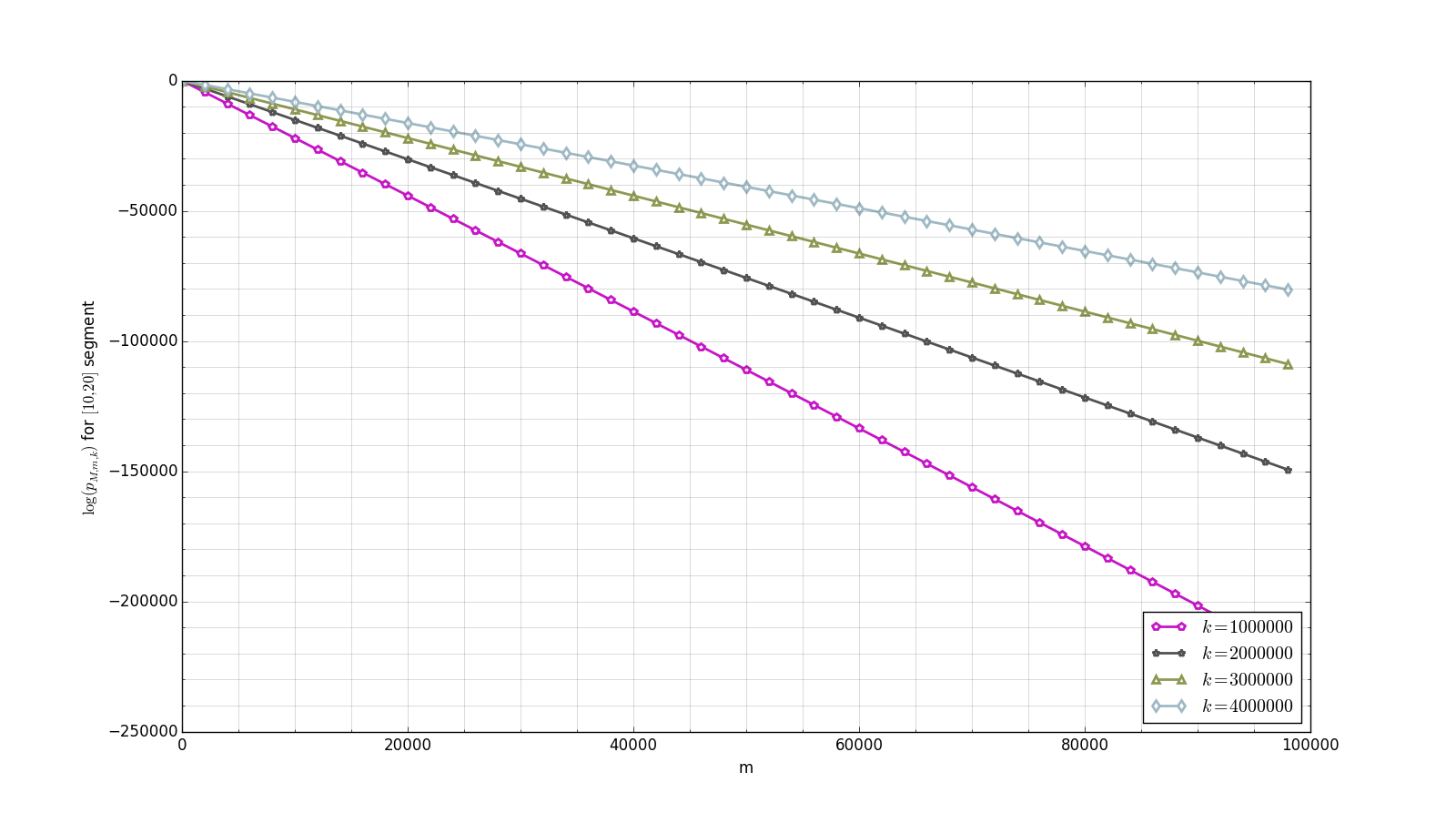}
        \caption{$\log(p_{M,m,k})(m)$ for $[10, 20]$ segment, $k \in \{10^6, 2\cdot 10^6, 3 \cdot 10^6, 4 \cdot 10^6\}$}
        \label{fig:logp-m-1}
    \end{subfigure}
\   ~
    \begin{subfigure}[b]{0.4\textwidth}
        \includegraphics[width=\textwidth]{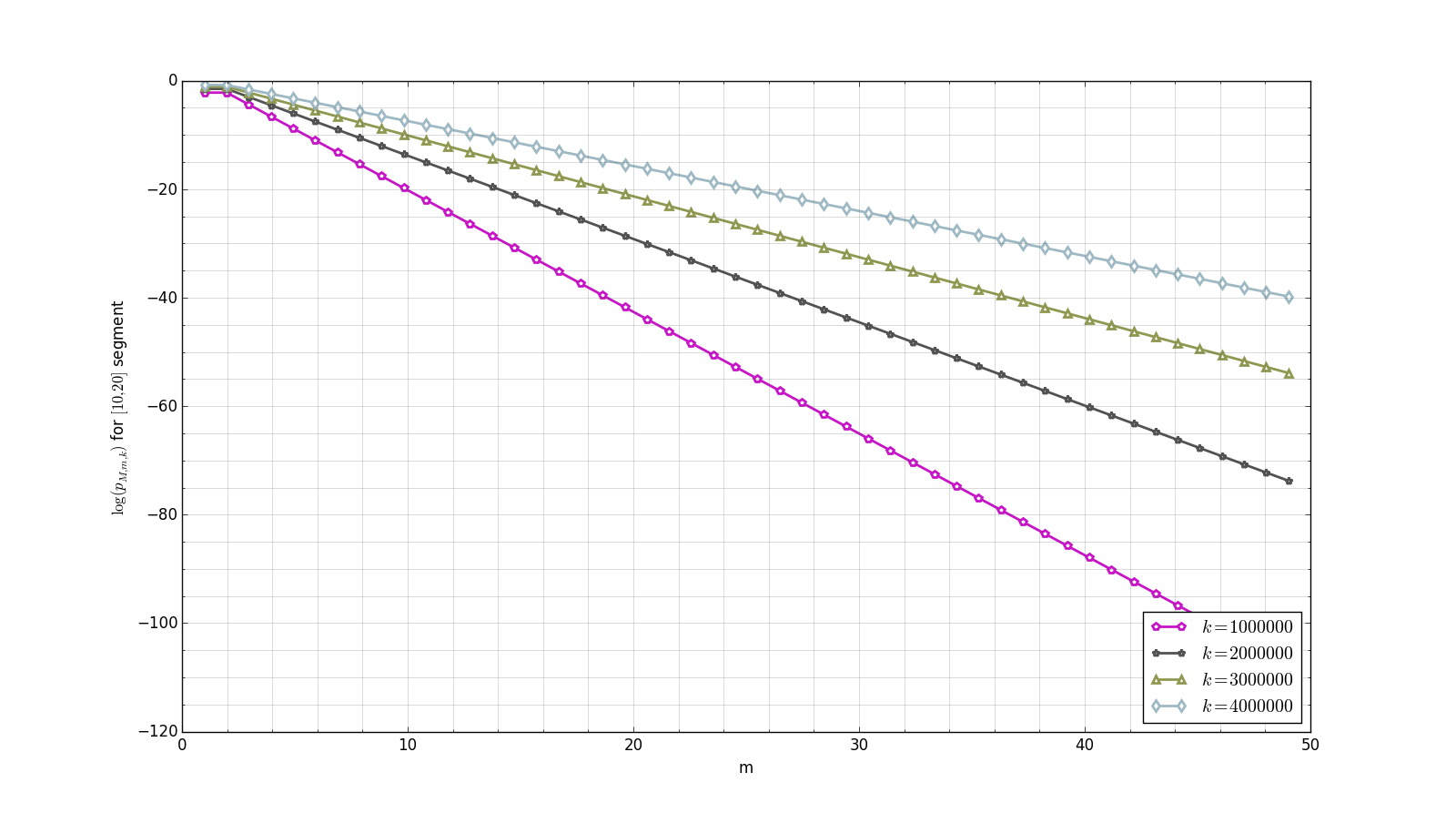}
        \caption{$\log(p_{M,m,k})(m)$ for $[10, 20]$ segment, $k \in \{10^6, 2\cdot 10^6, 3 \cdot 10^6, 4 \cdot 10^6\}$}
        \label{fig:logp-m-2}
    \end{subfigure}
    \caption{$\log(p_{M,m,k})$ with respect to $k$ (a, b) and $m$ (c, d).}
    \label{fig:logp}
\end{figure}

The graph in \figref{fig:pMmk-10e9} shows the behavior of $ p_{M,m,k} $ with respect to $ k $
for different values of $ m $.
The $ [10^9, 10^9 + 1] $ segment was taken,
and an appropriate estimate for the quantity of floating point numbers
is $ M = 9007199 $ in this case.
As can be seen in \figref{fig:M-wrt-A}, the value of $ M $ is very high
for small values of segment bounds. Such values make the computation of $ p_{M,m,k} $
quite complicated, since the intermediate values become too low
and will be soon treated as zeros.
So we take a segment of small length that is far from zero in order
to retrieve computable values of the error probability.
In order to illustrate more reasonable segments,
we consider $ \log(p_{M,m,k}) $.
One may notice that \eqref{eq:choose-m-in-a-row} implies
$ \log(p_{M,m,k}) = \log(\prod\limits_{i=0}^{m-1} \frac{k-i}{M-i})
= \sum\limits_{i=0}^{m-1}\log(\frac{k-i}{M-i})$.
Some plots of $ \log(p_{M,m,k}) $ are provided in \figref{fig:logp}.

Both \figref{fig:pMmk-10e9} and \figref{fig:logp} show that the values
of the error probability $ p_{M,m,k} $ are quite low for even huge number of zeros
of the target function and a relatively small amount of check points.
For instance, \figref{fig:logp-10-20} illustrates that, given a function with up to $ 10^6 $ zeros
on the segment $ [A,B] = [10, 20] $,
the probability of \algref{alg:add-compare-pseudo} error when checking up to $10$
points is lower than $ e^{-200} $.
Namely, so it is sufficient to check only $ 10 $ points on $ [10, 20] $
for such functions to be quite sure the result is correct.
It should also be noted that the way
in which $ \log(p_{M,m,k}) $ decreases as a function of $ m $
(\figref{fig:logp-m-1} and \figref{fig:logp-m-2})
seems to be linear (in fact, this is not the case),
which means that, for fixed values of $ M $ and $ k $,
the increase in $ m $ leads to nearly exponential decrease
in the error probability (in fact, it falls even faster).

\subsection*{Results}

The results of the error probability analysis show that this simple algorithm
can be quite successfully used to provide additional answer checking
in the Interactive Educational System.
It will not be too long for the server to check even up to $1000$ points on some
segment, and that leads to extremely low error probabilities for almost any
estimates of the number of zeros of an analytic function.
In fact, the only case,
where the quantity of zeros of such a function on a closed segment
is somehow comparable with the amount of floating point numbers in that segment
(so the error probability grows higher),
is the situation where this function has equilibrium points.
And the failure is doubtful even in that case.
But still to avoid an accidental case where an equilibrium point appears
inside or quite close to the testing segment,
it makes sense to run the algorithm
several times for several random disjoint segments.

It is important to understand that each function
we test by the proposed algorithm is always a difference of two functions:
the correct answer for the problem and the one obtained by the user.
Its nonzero value means that the user had obtained a wrong result.
The origins of such errors are some misunderstandings or accidental mistakes.
For instance, one may forget the minus sign when calculating the derivative
of $ \cos(x) $, and some arithmetic mistakes are also quite common.
The point is that big values of $ k $, when in fact user's result is wrong,
would mean that the two answers are really very close but still not equal.
And we believe that it is impossible for the user to obtain a wrong result
that would be so close to the correct one that the probability
of the proposed algorithm error is not extremely low.
For example,
as \figref{fig:logp}d shows, when checking up to $ 10 $ points inside $ [10, 20] $,
the two answers must concur in at least $ 4 \cdot 10^6 $ points
so that the error probability would be as high as $ \sim 10^{-5} $.
Compare this with the following situation.
Consider verbal examination with question cards.
The usual quantity of such cards is some multiple of $ 10 $.
Now consider the student who had prepared an excellent answer to the only one card,
while he cannot say a word answering to a question from any other card.
The probability that the student will successfully pass this exam is $ \sim 10^{-1} $,
however such a kind of examination is considered a good knowledge-checking practice.

It is rather strange that it has never been mentioned
that there is no way to guarantee correct answer checking
in educational systems which deal with human-defined mathematical expressions.
Although the problem seldom occurs in real-life examples,
it is important to know that it exists
and to understand how to behave in this case
and what the result reliability is.
We believe that the results obtained are going to help to avoid some sad effects
of the constant problem on educational systems with automatic answer checking development.

\end{document}